\documentclass[aps,prb,preprint,groupedaddress,showpacs,amsmath,amssymb]{revtex4}

\usepackage{graphicx}
\usepackage{dcolumn}
\usepackage{bm}

\begin{document}

\title{Effects of Co substitution on thermodynamic and transport properties and anisotropic $H_{c2}$ in Ba(Fe$_{1-x}$Co$_x$)$_2$As$_2$ single crystals.}

\author{N. Ni,$^1$ M. E. Tillman,$^1$ J.-Q. Yan,$^1$ A. Kracher,$^1$ S. T. Hannahs,$^2$ S. L. Bud'ko,$^1$ and P. C. Canfield$^1$}
\affiliation{$^1$Ames Laboratory US DOE and Department of Physics and Astronomy, Iowa State University, Ames, IA 50011, USA
\\
$^2$National High Magnetic Field Laboratory, 1800 East Paul Dirac Drive, Tallahassee, FL 32310, USA}

\date{\today}

\begin{abstract}

Single crystalline samples of Ba(Fe$_{1-x}$Co$_x$)$_2$As$_2$ with $x < 0.12$ have been grown and characterized via microscopic, thermodynamic and transport measurements.  With increasing Co substitution, the thermodynamic and transport signatures of the structural (high temperature tetragonal to low temperature orthorhombic) and magnetic (high temperature non magnetic to low temperature antiferromagnetic) transitions are suppressed at a rate of roughly 15 K per percent Co.  In addition, for $x \ge 0.038$ superconductivity is stabilized, rising to a maximum $T_c$ of approximately 23 K for $x \approx 0.07$ and decreasing for higher $x$ values.  The $T - x$ phase diagram for Ba(Fe$_{1-x}$Co$_x$)$_2$As$_2$ indicates that either superconductivity can exist in both low temperature crystallographic phases or that there is a structural phase separation.  Anisotropic, superconducting, upper critical field data ($H_{c2}(T)$) show a significant and clear change in anisotropy between samples that have higher temperature structural phase transitions and those that do not.  These data show that the superconductivity is sensitive to the suppression of the higher temperature phase transition.

\end{abstract}


\maketitle

\section{Introduction}

The discovery of superconductivity in F-doped LaFeAsO \cite{kam08a} and K-doped BaFe$_2$As$_2$ \cite{rot08a} compounds  has lead to an extensive experimental effort to characterize and delineate the nature of superconductivity in RFeAsO and (AE)Fe$_2$As$_2$ materials (R = rare earth; AE = Ba, Sr, Ca).  Currently the availability of single crystalline samples of significant size and quality \cite{nin08a,wan08a,yan08a,nin08b,ron08a,wug08a,sef08a} has made the study of the doped (AE)Fe$_2$As$_2$ compounds more tractable.  This, combined with the realization that Co substitution for Fe could be used to both stabilize superconductivity \cite{sef08a,lei08a} and simplify growth while improving homogeneity, makes the systematic study of the thermodynamic and transport properties of the Ba(Fe$_{1-x}$Co$_x$)$_2$As$_2$ series particularly compelling. So far only a limited set of data on the Ba(Fe$_{1-x}$Co$_x$)$_2$As$_2$ series: the effect of pressure on phase transitions in Ba(Fe$_{1-x}$Co$_x$)$_2$As$_2$ with $x = 0.04$ and 0.1, \cite{ahi08a} anisotropic $H_{c2}$ in Ba(Fe$_{0.9}$Co$_{0.1}$)$_2$As$_2$ \cite{yam08a} and details of superconducting state for Ba(Fe$_{0.93}$Co$_{0.07}$)$_2$As$_2$ \cite{pro08a,gor08a} have been reported.
\\

	In this work we combine detailed measurements of the low field thermodynamic and transport properties of single crystalline Ba(Fe$_{1-x}$Co$_x$)$_2$As$_2$ with high field measurements (up to 35 T) of the anisotropic upper superconducting critical field, $H_{c2}(T)$.  The analysis of these data allows us to create a $T - x$ phase diagram and shows that the superconductivity apparently occurs in both the low temperature orthorhombic and tetragonal phases, an observation that is clearly reflected in significant differences in the anisotropy of $H_{c2}$ in these two regions.

\section{Experimental Methods}

Single crystals of Ba(Fe$_{1-x}$Co$_x$)$_2$As$_2$ were grown out of self flux using conventional high-temperature solution growth techniques. \cite{sef08a,can92a} Small Ba chunks, FeAs powder and CoAs powder were mixed together according to the ratio Ba~:~FeAs~:~CoAs = $1:4(1-x):4x$. The mixture was placed into an alumina crucible. A second, catch crucible containing quartz wool was placed on top of this growth crucible and both were sealed in a quartz tube under $\sim 1/3$ atmosphere Ar gas. The sealed quartz tube was heated up to $1180^\circ$ C, stayed at $1180^\circ$ C for 2 hours, and then cooled to $1000^\circ$ C over 36 hours. To avoid cracking of the crucible, a $50^\circ$ C/hour heating rate was used. Once the furnace reached $1000^\circ$ C, the excess of FeAs/CoAs was decanted from the plate-like single crystals. The inset of Fig.2 (below) shows a picture of a single crystal of Ba(Fe$_{0.926}$Co$_{0.074}$)$_2$As$_2$ ($x_{nominal} = 0.10$) against a mm scale. Unlike the crystals grown out of Sn flux, \cite{nin08a,yan08a,nin08b} it doesn't have clear $(100)$ facets, although the plate itself is perpendicular to the crystallographic $c$-axis. Single crystal dimension can go up to $12~\times 8 \times 0.8$ mm$^3$.

The FeAs and CoAs powders used as part of the self-flux were synthesized by reacting Fe or Co powder and As powder after they were mixed together and pressed into pellets. The pellets were sealed inside a quartz tube under 1/3 atmosphere Ar gas, slowly heated up to $500^\circ$ C, kept at $500^\circ$ C for 10 hours, and then slowly heated up to $900^\circ$ C, and then kept at $900^\circ$ C for the other 10 hours.

	Powder x-ray measurements, with Si standard, were performed on a Rigaku Miniflex diffractometer using Cu $K \alpha$ radiation at room temperature. The lattice parameters were refined by "UnitCell" software. \cite{hol97a} The error bars are taken as $\pm 2\sigma$. Elemental analysis of the samples was performed using wavelength dispersive x-ray spectroscopy (WDS) in the electron probe microanalyzer of a JEOL JXA-8200 Superprobe. Heat capacity data were collected using a Quantum Design (QD) Physical Property Measurement System (PPMS). Magnetization and temperature-dependent ac ($f = 16$ Hz, $I = 3$ mA) electrical transport data were collected by a QD Magnetic Property Measurement System (MPMS). Electrical contacts were made to the sample using Epotek H20E silver epoxy to attach Pt wires in a four-probe configuration. The typical samples for electrical transport had approximately $0.1 \times 0.5$ mm$^2$ cross-section and the distance between the voltage contacts $\sim 1$ mm.

	Whereas the single crystals could be cut and cleaved into well defined geometries, we found that the values of resistivity we inferred from these samples could vary significantly.  Figure 1a presents data from three samples of Ba(Fe$_{0.926}$Co$_{0.074}$)$_2$As$_2$.  There is a clear difference between the three  $\rho (T)$ plots, but the basic features are also very similar, as well as the characteristic temperatures.  If we assume that cracks or exfoliations internal to the sample are responsible for these differences (effectively leading to errors in the geometric factor), then normalizing the room temperature slopes, $d \rho /d T|_{300 K}$, should compensate for this.  Figure 1b confirms this assumption, and Fig. 1c shows that allowing for slightly different residual resistivity scattering,  $\rho_0$, collapses the three curves onto each other.  Unfortunately as we change the amount of Co substitution we cannot use such a simple normalization since $d \rho/dT$ may easily be dependent on Co substitution levels.  For this reason we will present our data in terms of  $\rho (T)/\rho (300 K)$ (or equivalently $R(T)/R(300 K)$) when we compare different Co substitution levels.

	Field-dependent DC electrical transport data were collected using the 35 T and 33 T resistive magnets in the National High Magnetic Field Laboratory (NHMFL) in Tallahassee, FL for superconducting members of the series.  Two samples with each nominal concentration of Co were run, each for $H \| c$ and $H$ parallel to the basal plane (shown throughout the text as $H \perp c$).  Each of these samples was measured in zero field in the QD PPMS and MPMS units and the zero field $T_c$ was used to correct for slight temperature off-sets associated with the resistive probe used at the NHMFL.  These shifts were systematic (associated with the sample position relative to the thermometer) and were at most $10\%$ of $T_c$ and, for most runs, significantly smaller.

\section{Results}

In order to present the data in terms of actual $x$ rather than $x_{nominal}$,  elemental analysis was performed using wavelength dispersive x-ray spectroscopy (WDS).  The average inferred $x$ value was determined by averaging the x value measured at several locations on the sample.  For example, the cobalt substitution level for the $x_{nominal} = 0.10$ was measured to be 0.072, 0.070, 0.073 on three different layers of a crystal from one batch and 0.076, 0.076, 0.075 on three different layers of a crystal from a different batch, leading to an average, inferred substitution level of 0.074.  As shown in Fig. 2, there is a finite, but limited spread of measured Co concentrations.  In addition, there is a well defined, essentially linear, relation between nominal and actual Co concentration in the samples over the range of substitution we studied ($x_{WDS} = 0.74~ x_{nominal}$).  Figure 2 also shows that the unit cell volume follows an essentially linear dependence on Co concentration.  To this end, the unit cell volume can be used to estimate the Co doping levels from simple powder x-ray diffraction.  For this work, the inferred Co values will be used to identify the samples.

The unit cell volume data shown in Fig. 2 was determined via powder x-ray diffraction patterns taken on ground single crystals from each batch. No impurity phases could be detected and the tetragonal unit cell parameters, $a$ and $c$, could be readily inferred.  These are plotted in Fig. 3 as relative changes in lattice parameter.  There is a clear, monotonic, and essentially linear decrease in the $a$- and $c$-lattice parameters as well as the unit cell volume up to $x \sim 0.12$.

	The effects of Co substitution on the low temperature properties of Ba(Fe$_{1-x}$Co$_x$)$_2$As$_2$ crystals can be clearly seen in transport data (Fig. 4).  For BaFe$_2$As$_2$, $x = 0.00$, the sharp decrease in resistivity associated with the combined structural and antiferromagnetic transitions \cite{rot08b,hua08a} is clearly seen at 134 K.  As $x$ is increased the temperature of the resistive anomaly is suppressed monotonically, no longer being detectable for $x > 0.058$.  For $x \ge 0.038$ superconductivity is readily detected by a sharp (and complete) decrease of the resistivity to zero.  This is shown more clearly in the inset to Fig. 4.  For $x = 0.038, 0.047$ and even 0.058 both the higher temperature, resistive anomaly and superconductivity are detected.  It is worth noting that even our lowest Co substitution level, $x = 0.013$, causes a complete change in sign of the resistive feature; the resistivity increases below the high temperature transition rather than decreasing.  Whereas for $x = 0.013$ and 0.020 the resistive anomaly is very reminiscent of that seen for pure CaFe$_2$As$_2$, \cite{nin08b,ron08a,wug08a} for $x = 0.038$ and 0.047, the resistive anomaly is very similar to that originally found for BaFe$_2$As$_2$ grown out of Sn flux \cite{nin08a} (i.e. when $\sim 1 - 2 \%$ Sn was found to substitute for the As).  A similar change in the sign of the resistivity anomaly can be seen by comparing the resistivity for single crystal SrFe$_2$As$_2$ \cite{yan08a} to that for Sr(Fe$_{1-x}$Co$_x$)$_2$As$_2$. \cite{lei08a} For the smallest reported Co doping level in Sr(Fe$_{1-x}$Co$_x$)$_2$As$_2$ an increase similar to what we find for $x = 0.038$ and 0.047 in Ba(Fe$_{1-x}$Co$_x$)$_2$As$_2$ can be found.

Both the antiferromagnetic/structural phase transition as well as the superconductivity can also be detected in temperature dependent magnetization measurements.  Fig. 5 presents the $M/H$ data for $H = 1$ T applied perpendicular to the crystallographic $c$-axis of the samples.  For $0.00 \le x \le 0.047$ (upper panel) there is a clear drop in the susceptibility at the temperature associated with the resistive anomalies (as well as a  lower temperature feature associated with superconductivity for $x = 0.038$ and 0.047).  For larger $x$ values (Fig. 5 lower panel) superconductivity can be observed, but there is no longer any signature of the higher temperature transition.   The superconducting transition can be seen more clearly in the low field $M/H$ data shown in Fig. 6.  Each data set has been converted into units of $1/4 \pi$  by using the unit cell volume and inferred Co doping level from the data presented in Fig. 3.  The upper curves are field cooled data whereas the lower curves are zero-field-cooled warming data.

	Specific heat data ($H = 0$ and 9 T applied along the crystallographic c-axis) was measured for Co-doping levels of $x = 0.038$ and 0.047 (Fig. 7) as well as for $x = 0.078$ and 0.10 (Fig. 8).  For $x = 0.038$ two breaks in slope can be seen at higher temperatures.  These can be more clearly identified by examining the $dC_p/dT$ plot in the lower inset.  A similar pair of features can also be seen for the $x = 0.047$  data.  A lower temperature feature in the specific heat can be clearly seen near the $T_c$ values identified from the resistivity and susceptibility data shown in Figs. 4-6.  These features are even more clearly resolved by comparison with the 9 T data.  Figure 8 presents the low temperature $C_p(T)$ data for $x = 0.078$ and 0.10 samples. For $x = 0.078$ (Fig. 8, upper panel) the specific heat anomaly is relatively sharp and readily associated with a second order phase transition to a low temperature superconducting state.  For the other $x$ values, low temperature, anomalies associated with the superconducting phase transition are broader and will be discussed in detail below.

	The resistivity, susceptibility and specific heat data all clearly show the suppression of the higher temperature phase transition (associated with a structural and antiferromagnetic phase transition in BaFe$_2$As$_2$ \cite{nin08a,rot08b,hua08a}) with increasing Co substitution for Fe.  In order to quantify this more fully we have to introduce criterion for determining the salient transition temperatures.  For $x \le 0.020$ a single transition is clearly identifiable in both resistivity and magnetization, but for higher $x$ values the transition appear to broaden and possibly even split.  Figures 9 shows the temperature derivatives of the specific heat, resistivity and susceptibility for $x = 0.038$.  Clear features in the resistivity and susceptibility occur at the similar temperatures as the breaks in slope in the specific heat.  (Similar features in the temperature derivatives of resistivity at ambient and hydrostatic pressure were observed  in Refs. \onlinecite{ahi08a,tor08a} as well.) Transition temperature data inferred in this manner are plotted as a $T - x$ phase diagram in Fig. 10.  In a similar manner criteria for the determination of $T_c$ have to be established and used.  For this study we use:  (i) an offset criterion for resistivity (the temperature at which the maximum slope of the resistivity curve extrapolates to zero resistance); (ii) an onset criterion for susceptibility (the temperature at which the maximum slope of the susceptibility extrapolates to the normal state susceptibility); and (iii) the temperature at which the zero field specific heat deviates from the 9 T specific heat.  The data points inferred from these criteria are shown in Fig. 10 as well.

	Figure 10 presents a summary of the microscopic, thermodynamic and transport data presented so far in the form of a temperature - doping phase diagram.  The higher temperature phase transition that is associated with a structural/antiferromagnetic transition is monotonically suppressed with increasing Co substitution and is no longer detected in either resistivity or magnetization data for $x > 0.058$.  Superconductivity is stabilized for $x = 0.038$, rises to a maximum value ($T_c \sim 23$ K) for $0.058 \le x \le 0.074$, and drops to $T_c \sim 10$ K again by $x = 0.11$.  Whereas the resistive and, perhaps more importantly, the magnetic signatures of superconductivity are sharp, the feature associated with superconductivity in specific heat was only sharp for the sample with near maximum $T_c$.  It is important to note that the transition temperatures inferred from resistivity data (squares), magnetization data (circles) and specific heat data (triangles) are in excellent agreement with each other.

	One of the clearest features in Fig. 10, as well as the resistivity and magnetization data shown in Figs. 4 and 5 is the apparent coexistence of the higher temperature, structural (antiferromagnetic) phase transition with lower temperature superconductivity for the lower Co-doping levels.  This could imply that superconductivity does exist in the low temperature, orthorhombic (and presumably antiferromagnetic) state.  This, of course presumes that the same parts of the sample are simultaneously superconducting and orthorhombic.  Another possibility would be that the sample separates into tetragonal (superconducting) and orthorhombic (not superconducting) phases.

	In order to study the effects of doping on the superconductivity in greater detail, anisotropic $H_{c2}(T)$ data were collected for two samples for each doping level that manifested superconductivity.  A representative data set of $R(H)$ data for $H \| c$ and $H \perp c$ and $x = 0.038$ is shown in Fig. 11.  The two criteria that were used to infer $H_{c2}$ from a given $R(H)$ curve:  onset and offset are shown in Fig. 11.  These two criteria are representative of the width of the transition seen in the $R(H)$ data and will give a sense of the range of possible $H_{c2}(T)$ values.  Figures 12 - 17 present the anisotropic $H_{c2}(T)$ plots inferred from the $R(H)$ data for each criterion.  The $H_{c2}(T)$ data manifest several trends, some trivial, some more profound.   In the Ba(Fe$_{1-x}$Co$_x$)$_2$As$_2$ series, grossly speaking, the $H_{c2}(T)$ curves become higher for higher $T_c$ values.  Whereas, for the samples with lower $T_c$ values the 35 T maximum field was enough to fully determine the $H_{c2}(T)$ plots, the samples with $T_c$ values above 20 K have $H_{c2}(T=0)$ values that are significantly larger. The $H_{c2}$ anisotropy appears to be significantly smaller than that in superconducting oxypnictides, \cite{mar08a,wel08a,bal08a,jar08a} but  for all values of $x$, the $H_{c2}(T)$ curve for $H \perp c$ is higher than that for $H \| c$.  In addition, for all values of $x$, close to $T_c$, the upper, $H \perp c$ curve is closer to linear than the $H \| c$ curve which manifests an increasing (negative) slope as $T$ is reduced. The close to linear $H_{c2}(T)$ not far from $T_c$ is reminiscent to the standard WHH \cite{wer66a} behavior, whereas a positive curvature of $H_{c2}(T)$ is sometimes considered a hallmark of multiband superconductivity. \cite{shu98a} These differences in $H_{c2}(T)$ behavior may possibly be accounted for by complex multi-sheet Fermi surfaces of these materials, but detailed, realistic, modeling based on band-structure calculations are required to address  this issue. The sign and the order of magnitude of the $H_{c2}$ anisotropy, as well as difference in curvature of $H_{c2}(T)$ for two different orientations are similar to that observed in K-doped BaFe$_2$As$_2$ \cite{alt08a,yua08a} as well as to data taken on nominal $10\%$ Co-doped BaFe$_2$As$_2$. \cite{yam08a}

\section{Discussion}

The temperature - composition phase diagram shown in Fig. 10 summarizes the features of the Ba(Fe$_{1-x}$Co$_x$)$_2$As$_2$ series:  substitution of Co for Fe suppresses the high temperature, tetragonal-to-orthorhombic (as well as paramagnetic to antiferromagnetic) phase transition, completely suppressing it for $x > 0.06$.  In addition for $x \ge 0.038$, and up through our highest doping level of $x = 0.114$, superconductivity can be stabilized, with a broad maximum, with $T_c \sim 23$ K for $x \sim 0.07$. (A grossly similar phase diagram with superconductivity possibly emerging still in orthorhombic/magnetic phase was reported for Ba$_{1-x}$K$_x$Fe$_2$As$_2$, \cite{rot08c,che08a} with no evidence, though, for split structural/magnetic  transition.) Several questions about the details and implications of this phase diagram immediately arise:  what can be inferred from the broadened (or split) transitions found for intermediate Co concentrations and what can be inferred about the coexistence of superconductivity and the orthorhombic (antiferromagnetic) phase?

	The observation of two clear features in $d \rho/dT$ as well as $d\chi/dT$, coupled with features in $C_p$ for intermediate Co concentrations can be discussed within two very different frameworks:  a separation of the structural and magnetic phase transitions that are so strongly coupled for $x = 0$ (not only in BaFe$_2$As$_2$ but, for example, in CaFe$_2$As$_2$ \cite{gol08a}); or a distribution of transition temperatures associated with a distribution of Co doping throughout the sample.  The first hypothesis is certainly an exciting one but, unfortunately, the current data set cannot do more than address it peripherally.  If the two phase transitions (structural and magnetic) were to separate cleanly then there would be no reason for the broadened features in the resistivity as well as the somewhat subtle features in the specific heat.  If there indeed is a separation of the magnetic and structural phase transitions, then there also has to be some broadening of them as well (as inferred from these data).  The second hypothesis is somewhat more supported by the current data.  Figure 10 shows a roughly 15 K decrease in the upper transition temperature for each percent of Co substituted.  If we associate the two features we observe in the resistivity as well as in the susceptibility with a range of transition temperatures associated with the variation of Co content in the sample, then we can infer, that for increasing Co-substitution there may be a range of doping as large as $\pm 0.005$.  This is comparable (within a factor of two) to the variation in inferred WDS stoichiometry discussed in the experimental methods section.  This hypothesis also allows for an understanding of the initially puzzling variation of the specific heat feature associated with the superconducting transition.  Figures 7 and 8 indicate that for $x = 0.074$ there is a relatively sharp $C_p$ anomaly, but for other $x$ the feature is broadened.  If there is a distribution of Co concentrations throughout the sample, then this distribution will lead to a distribution of $T_c$ values that can be evaluated by looking at the local curvature of the $T_c(x)$ dome.  For the low $x$, or high $x$, side of the dome there is a clear slope and this will lead to a relatively large distribution of $T_c$ values.  On the other hand, for $x$ values near the optimal $x$ value (the one giving the highest $T_c$) a distribution of $x$ values will lead to a negligible spread in $T_c$.  This is consistent with what we can infer from the $C_p(T)$ data.

	The $T - x$ phase diagram presented in Fig. 10 raises another question concerning the homogeneity of the sample: if at $T_c$ the sample is assumed to be single phased (i.e. completely in the orthorhombic (antiferromagnetic) phase), then Fig. 10 clearly shows that superconductivity exists in either phase (orthorhombic and tetragonal) with apparent equal ease.  The alternate to this scenario is to assume that parts of the sample remain in the tetragonal state and these parts are the ones that become superconducting at low temperatures.  Some light can be shed on this hypothesis by examining how the properties of the superconducting state vary from one side of the dome to the other.  If the superconductivity is associated solely with the tetragonal phase, volume fraction of the sample, then there should not be a dramatic variation in the anisotropy of the superconductivity (all coming from the tetragonal phase).  On the other hand, if there is a clear difference in the superconducting anisotropy between the samples that are in the orthorhombic part of the phase diagram and the samples that are in the tetragonal part of the phase diagram, then this would be evidence of the superconductivity being affected by the higher temperature phase transition.  Figure 18 presents the anisotropy of $H_{c2}(T)$, $\gamma = H_{c2}^{\perp c}(T) / H_{c2}^{\|c}(T)$, plotted as a function of $T/T_c$.  The data for each of the two samples from each batch are shown and as well as the two different criteria.  There is a remarkable bifurcation of anisotropy, with the three Co doping values that are clearly in the tetragonal state having an anisotropy that is $40-60\%$ larger than the samples that manifest superconductivity as well as anomalies associated with the orthorhombic (antiferromagnetic) phase transition.  Although this is not proof that superconductivity can arise from the orthorhombic phase, it certainly is an evidence that the superconducting phase in the samples that do manifest signatures of this transition have a markedly different anisotropy.

\section{Summary and Conclusions}

	The effects of Co substitution for Fe has been studied in single crystals of Ba(Fe$_{1-x}$Co$_x$)$_2$As$_2$ for $x < 0.12$.  The addition of Co monotonically suppresses the higher temperature structural (antiferromagnetic) phase transition at an intial rate of roughly 15 K per percent Co.  In addition superconductivity is stabilized at low temperatures for $x \ge 0.038$ and up through our highest doping level of $x = 0.114$.  The superconducting region has a dome-like appearance with the maximum $T_c$ values ($\sim 23$ K) found near $x \sim 0.07$.  Although it is possible that intermediate values of Co doping lead to separation of the structural and magnetic phase transitions (as inferred by the observation of two features in the temperature derivatives of the thermodynamic and transport data) it is more likely that there is a finite (but small) spread in the local Co concentrations in different parts of the sample.  This hypothesis is supported by the observation that the sharpness of the $C_p$ anomaly associated with superconductivity tracking the curvature of the superconducting dome in the $T - x$ phase diagram.

	The $T - x$ phase diagram inferred for Ba(Fe$_{1-x}$Co$_x$)$_2$As$_2$ clearly shows the existence of superconductivity in the orthorhombic (antiferromagnetic) phase.  Although this may be a manifestation of the material breaking into superconducting (tetragonal) regions and non-superconducting (orthorhombic) regions, the analysis of the $H_{c2}$ anisotropy clearly shows that the superconductivity that occurs in samples that show features associated with the transition to the low temperature orthorhombic state has an $\sim 50 \%$ smaller anisotropy than that found in samples with higher Co-levels that remain in the tetragonal phase.  This observation is consistent with the superconductivity being affected by a different crystallographic environment.


\begin{acknowledgments}
Work at the Ames Laboratory was supported by the US Department of Energy - Basic Energy Sciences under Contract No. DE-AC02-07CH11358. A portion of this work was performed at the National High Magnetic Field Laboratory, which is supported by NSF Cooperative Agreement No. DMR-0084173, by the State of Florida, and by the DOE. We thank Kevin Dennis and R. William McCallum for useful discussions and experimental assistance.
\end{acknowledgments}

\clearpage

\begin{figure}
\begin{center}
\includegraphics[angle=0,width=120mm]{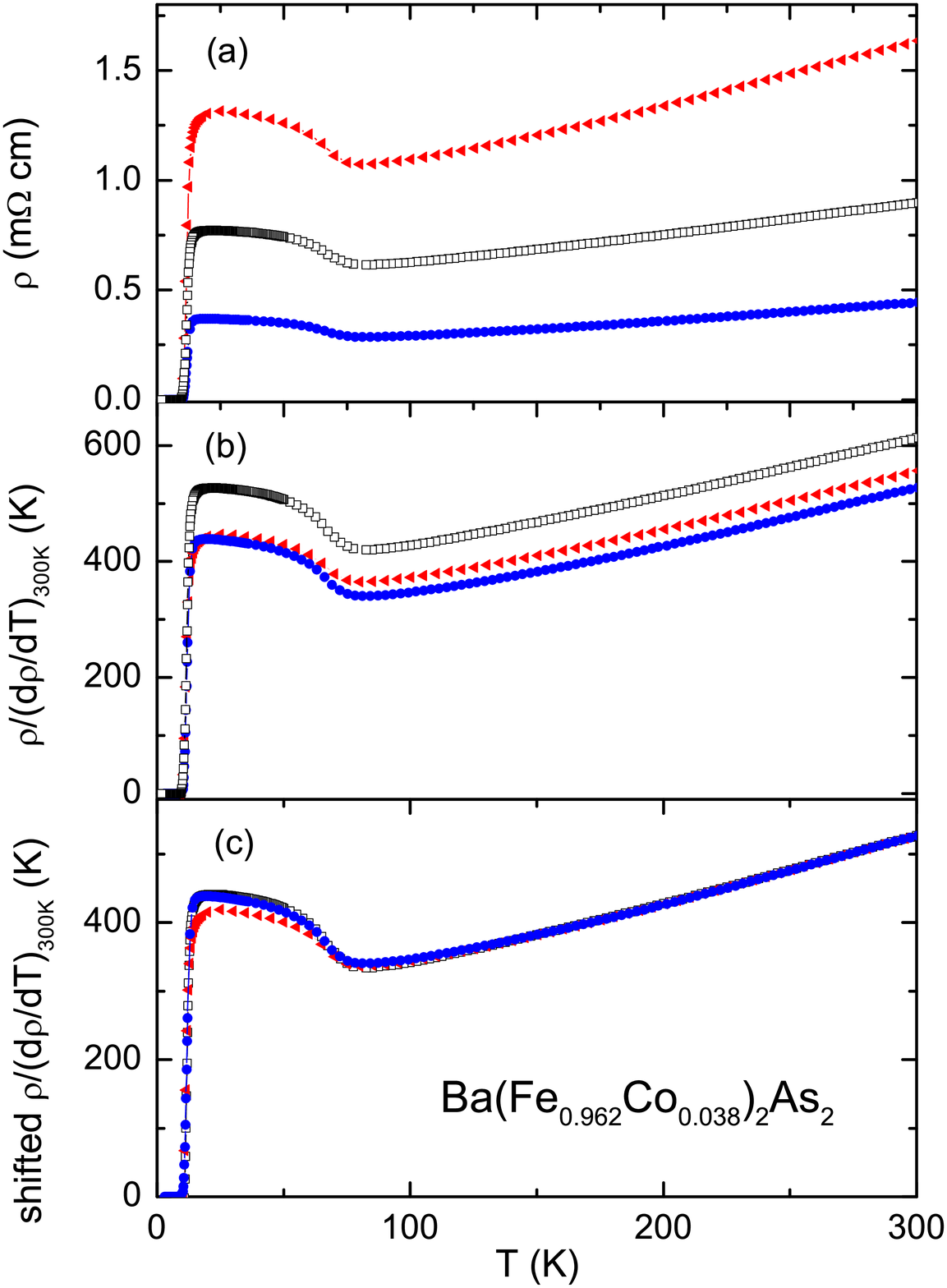}
\end{center}
\caption{(Color online) (a) Inferred temperature-dependent electrical resistivity of three samples of Ba(Fe$_{0.962}$Co$_{0.038}$)$_2$As$_2$. (b) Temperature-dependent electrical resistivity of three samples of Ba(Fe$_{0.962}$Co$_{0.038}$)$_2$As$_2$ normalized to their respective room temperature slopes: $\rho (T)/(d \rho /dT)|_{300 K}$. (c) Same data as (b) with upper curve shifted down by 85.7 K and intermediate curve shifted down by 28 K to account for differences in temperature independent, residual resistivity.}\label{F1}
\end{figure}

\clearpage

\begin{figure}
\begin{center}
\includegraphics[angle=0,width=120mm]{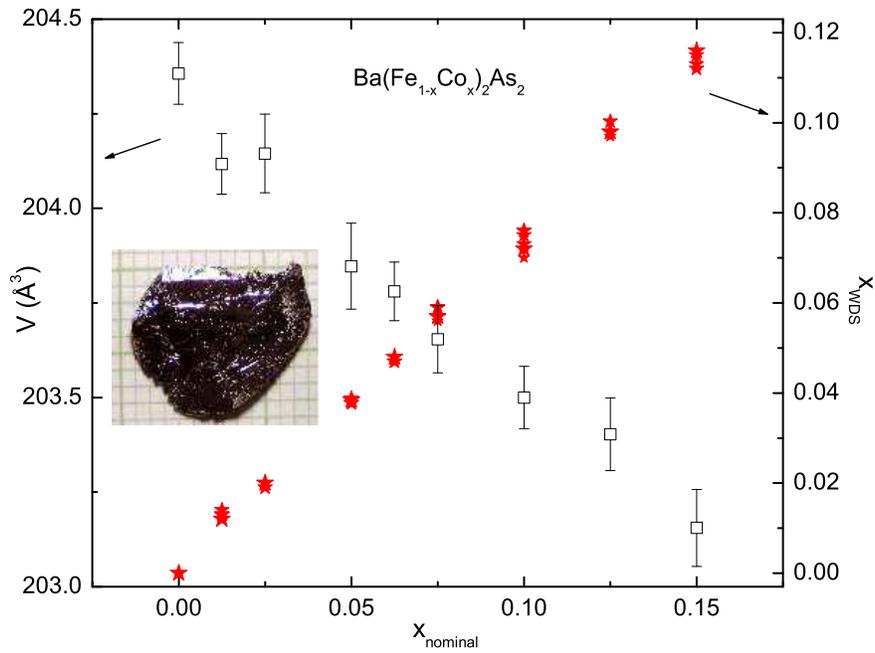}
\end{center}
\caption{(Color online) Unit cell volume and Co concentration determined from WDS measurement as a function of nominal Co concentration.  Multiple WDS data points were collected for each nominal $x$ and are each plotted, giving a sense of measured variation in Co concentration.  Inset:  picture of a representative single crystal over a mm grid.}\label{F2}
\end{figure}

\clearpage

\begin{figure}
\begin{center}
\includegraphics[angle=0,width=120mm]{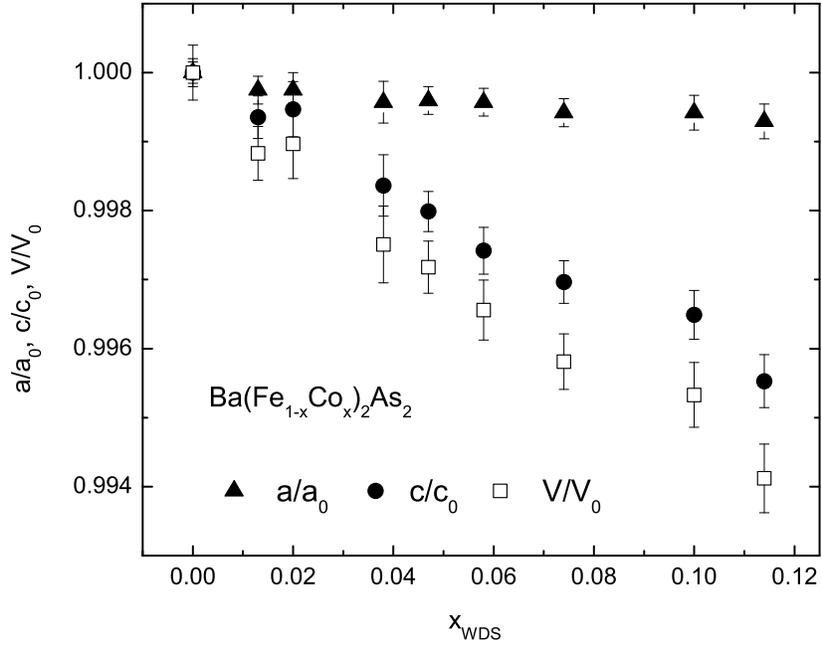}
\end{center}
\caption{Unit cell parameters, $a$ and $c$ as well as unit cell volume, $V$, normalized to $a_0 = 3.9621~\AA$, $c_0 = 13.0178~\AA$, $V_0 = 204.3565~\AA^3$ of undoped BaFe$_2$As$_2$ as a function of measured concentration of Co, $x_{WDS}$.}\label{F3}
\end{figure}

\clearpage

\begin{figure}
\begin{center}
\includegraphics[angle=0,width=120mm]{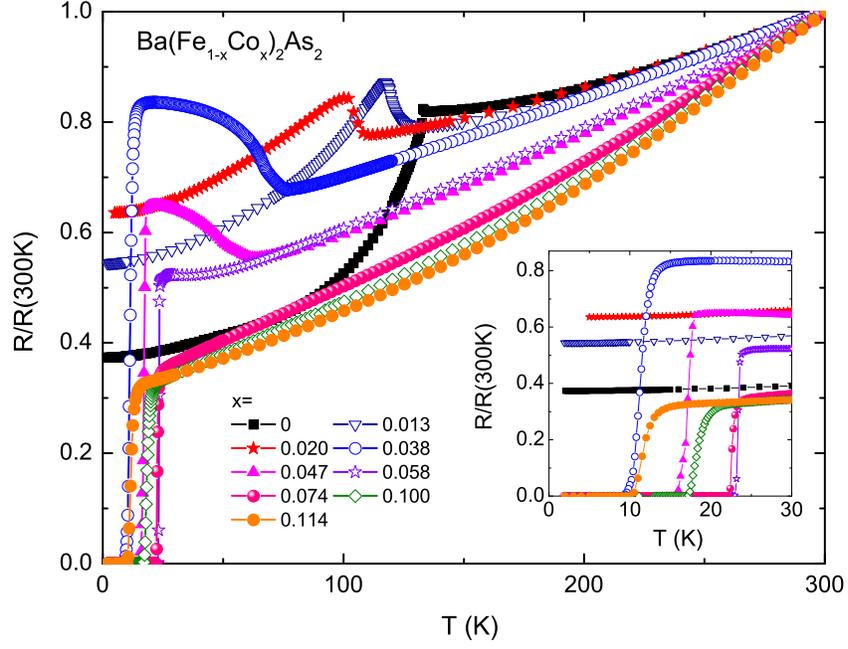}
\end{center}
\caption{(Color online) Electrical resistivity of Ba(Fe$_{1-x}$Co$_x$)$_2$As$_2$ single crystals normalized to their room temperature value (see text) for $0.00 \le x_{WDS} < 0.12$.  Inset:  Low temperature data showing superconducting transition for $x_{WDS} \ge 0.038$.}\label{F4}
\end{figure}

\clearpage

\begin{figure}
\begin{center}
\includegraphics[angle=0,width=120mm]{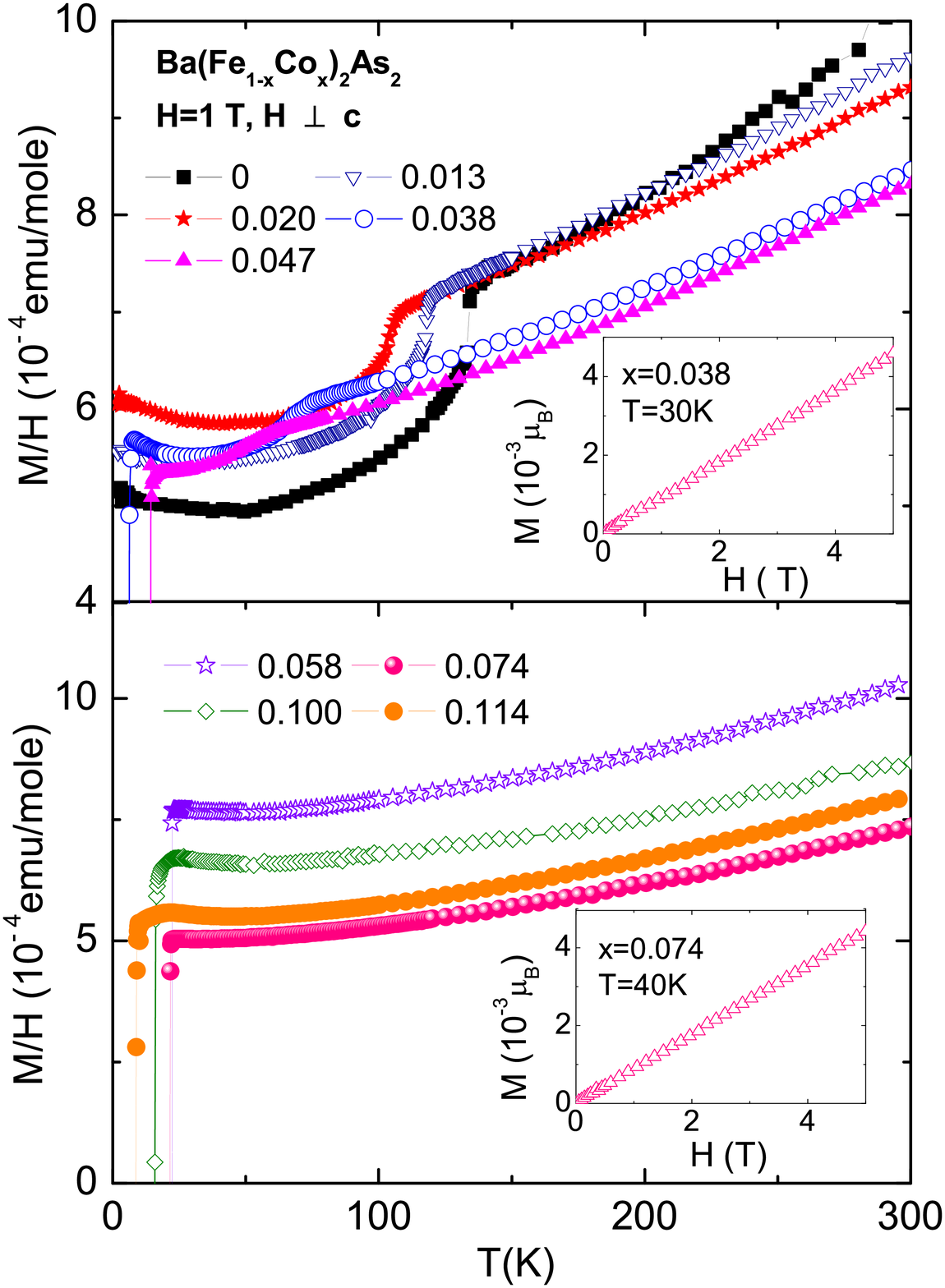}
\end{center}
\caption{(Color online) Magnetization divided by applied field, $M/H$, as a function of temperature for Ba(Fe$_{1-x}$Co$_x$)$_2$As$_2$ single crystals with a field of 1 T applied perpendicular to the crystallographic $c$-axis.   Upper panel shows data for $x_{WDS} \le 0.047$ and lower panel shows data for $x_{WDS} > 0.047$.  Insets show linearity of $M(H)$ data for selected $x_{WDS}$ values.
}\label{F5}
\end{figure}

\clearpage

\begin{figure}
\begin{center}
\includegraphics[angle=0,width=120mm]{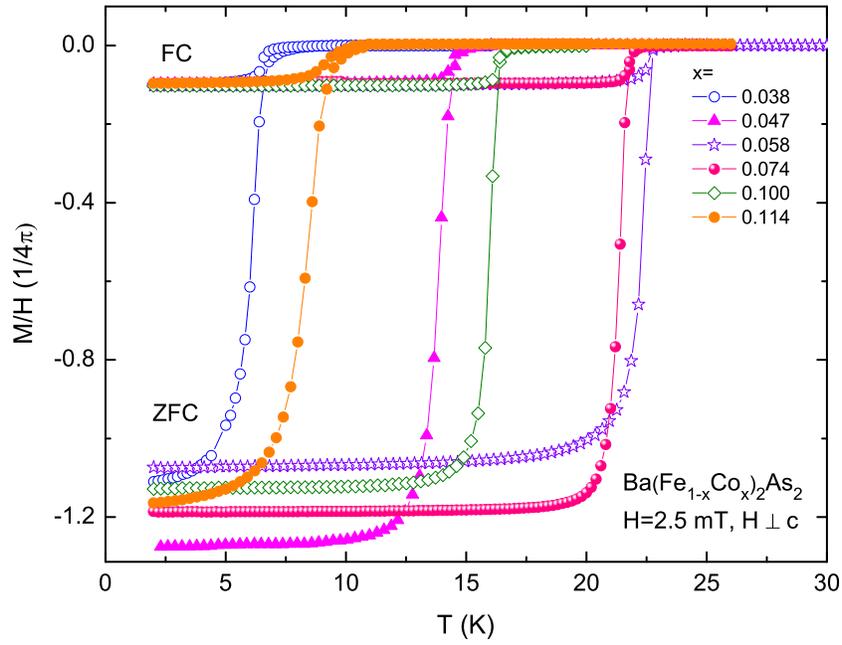}
\end{center}
\caption{(Color online) Magnetization divided by applied field, $M/H$, as a function of temperature for Ba(Fe$_{1-x}$Co$_x$)$_2$As$_2$ single crystals with a field of 2.5 mT applied perpendicular to the crystallographic $c$-axis.  Zero-field-cooled-warming data as well as field cooled data are shown.}\label{F6}
\end{figure}

\clearpage

\begin{figure}
\begin{center}
\includegraphics[angle=0,width=120mm]{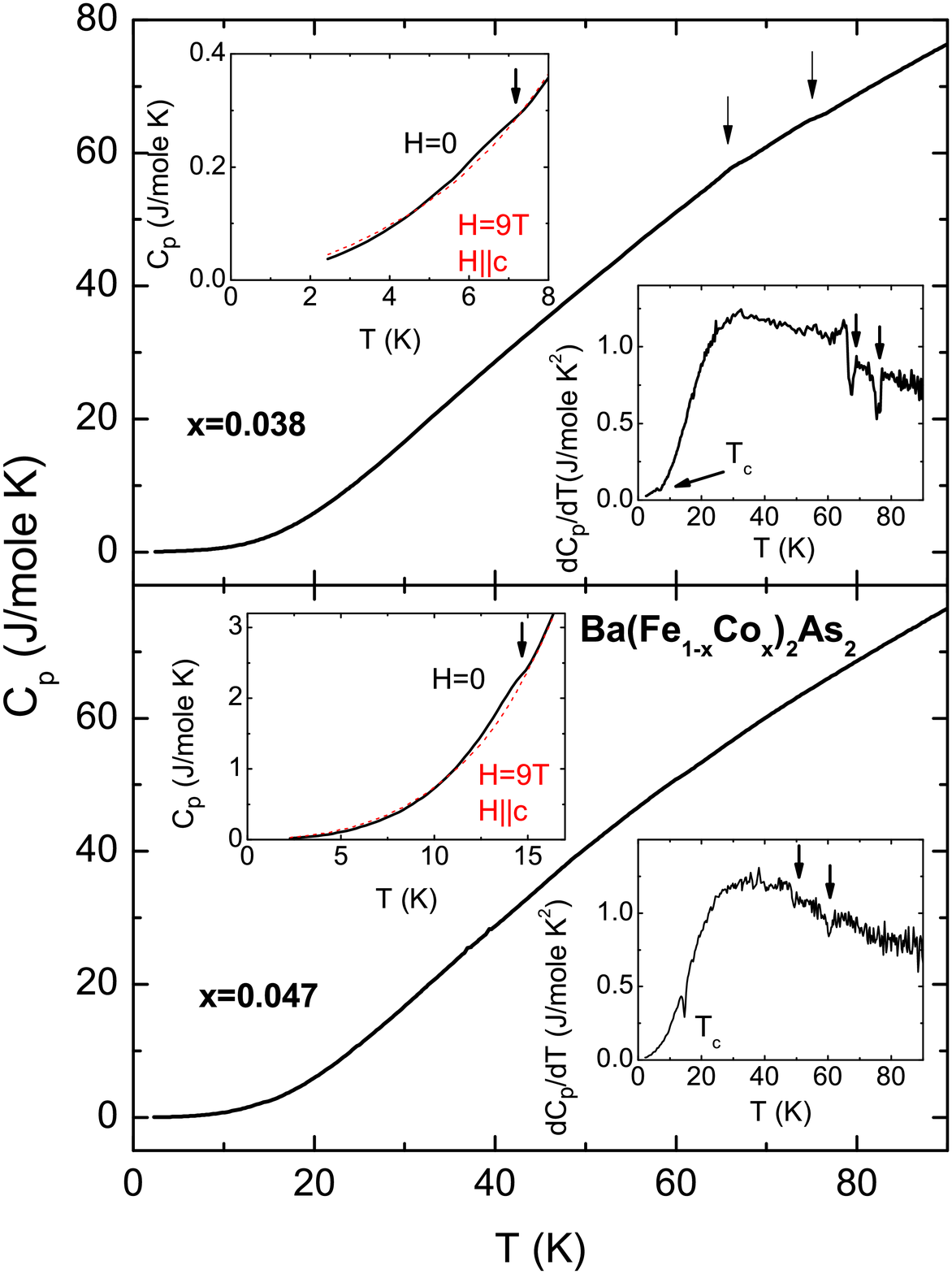}
\end{center}
\caption{ The temperature dependent specific heat of Ba(Fe$_{1-x}$Co$_x$)$_2$As$_2$ single crystals with $x  = 0.038$ (upper panel) and $x = 0.047$ (lower panel).   Lower insets:  $dC_p/dT$; upper insets:  Low temperature $C_p(T)$ data in zero (solid line) and 9 T (dashed line) applied along the crystallographic $c$-axis.}\label{F7}
\end{figure}

\clearpage

\begin{figure}
\begin{center}
\includegraphics[angle=0,width=120mm]{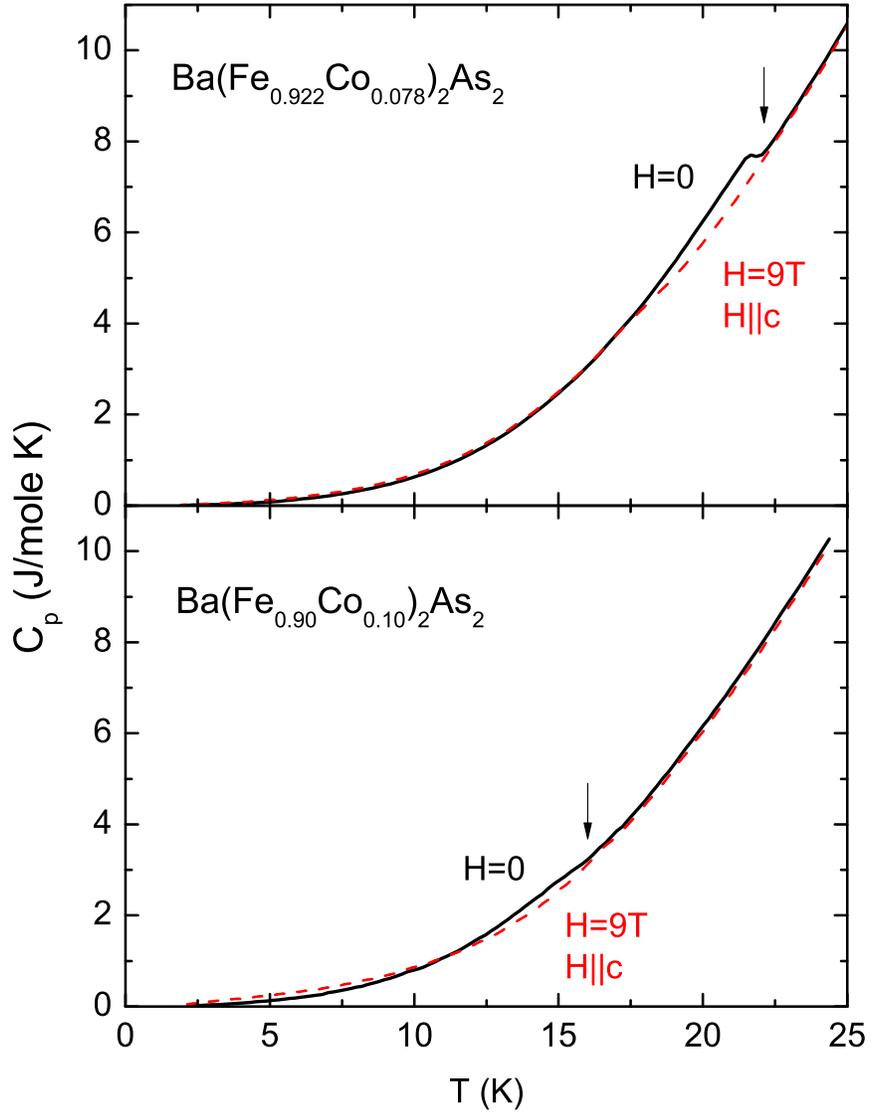}
\end{center}
\caption{The temperature dependent specific heat of Ba(Fe$_{1-x}$Co$_x$)$_2$As$_2$ single crystals with $x = 0.078$ (upper panel) and 0.10 (lower panel) in zero (solid line) and 9 T (dashed line) applied field along the crystallographic $c$-axis.}\label{F8}
\end{figure}

\clearpage

\begin{figure}
\begin{center}
\includegraphics[angle=0,width=120mm]{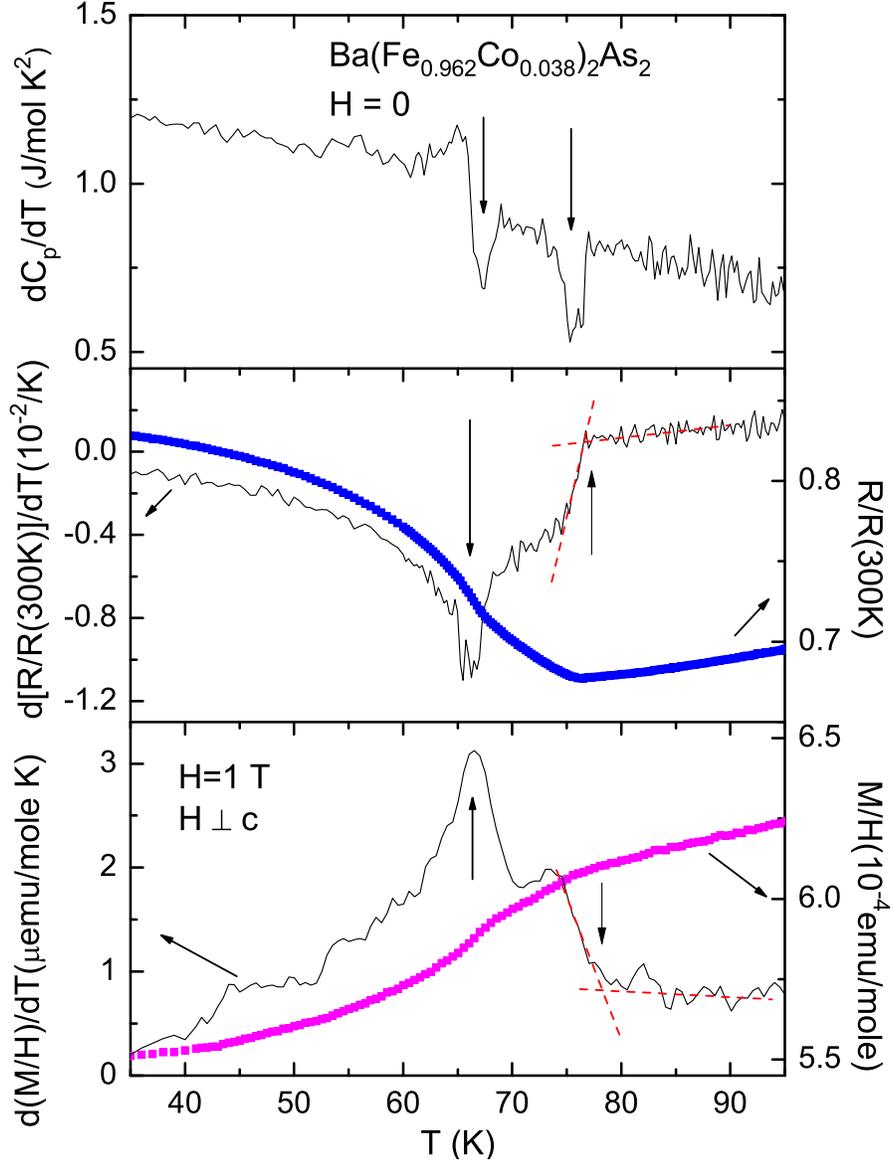}
\end{center}
\caption{(Color online) Criteria used to determine values for higher temperature transition(s).  Upper panel: $dC_p/dT$ emphasizes breaks in slope of $C_p(T)$ data.  Middle panel:  $(dR(T)/dT)/R(300 K)$ and $R(T)/R(300 K)$.  Bottom panel:  $d(M(T)/H)/dT$ and $M(T)/H$.}\label{F9}
\end{figure}

\clearpage

\begin{figure}
\begin{center}
\includegraphics[angle=0,width=120mm]{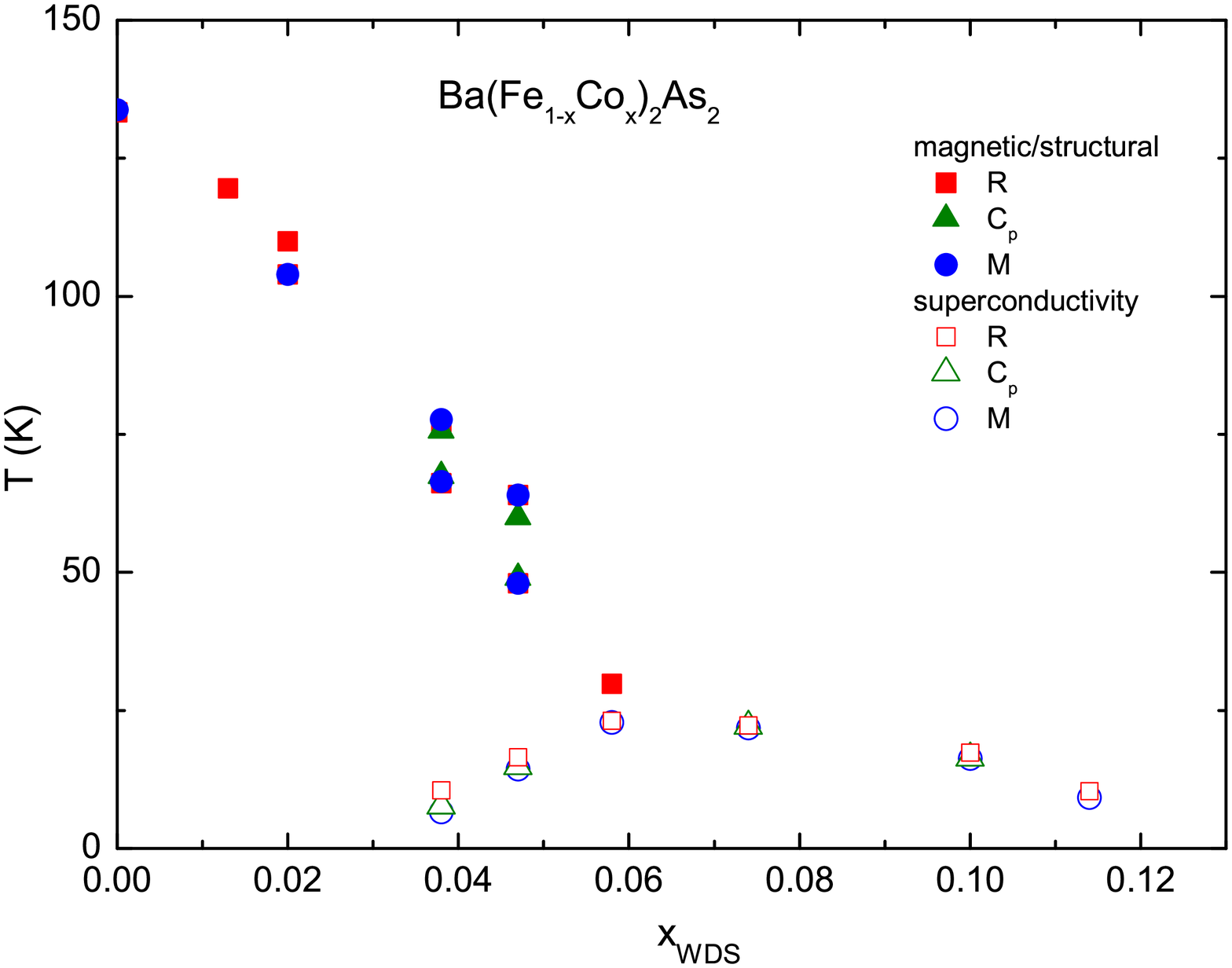}
\end{center}
\caption{(Color online) $T - x$ phase diagram for Ba(Fe$_{1-x}$Co$_x$)$_2$As$_2$ single crystals for $x < 0.12$.  Filled symbols represent transition temperatures associated with the higher temperature structural (antiferromagnetic) phase transition and open symbols represent transition temperatures associated with the superconducting phase transition.  Squares are data from resistivity data, circles are data from magnetization data and triangles are data from specific heat data.}\label{F10}
\end{figure}

\clearpage

\begin{figure}
\begin{center}
\includegraphics[angle=0,width=120mm]{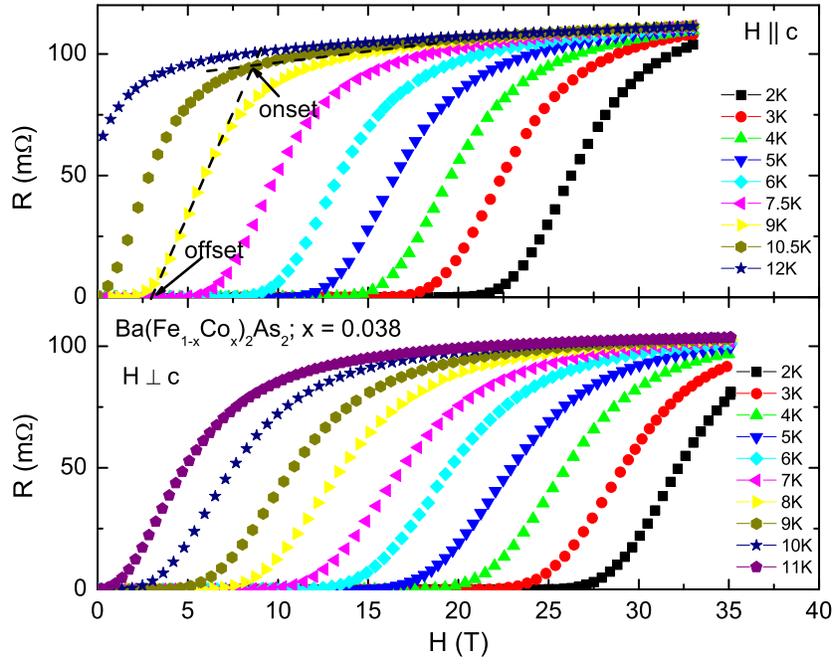}
\end{center}
\caption{(Color online) Isothermal $R(H)$ data from Ba(Fe$_{1-x}$Co$_x$)$_2$As$_2$ ($x = 0.038$) for $H \| c$ (upper panel) and $H \perp c$ (lower panel).  Dotted lines show onset and offset criteria used to determine $H_{c2}(T)$ values.}\label{F11}
\end{figure}

\clearpage

\begin{figure}
\begin{center}
\includegraphics[angle=0,width=120mm]{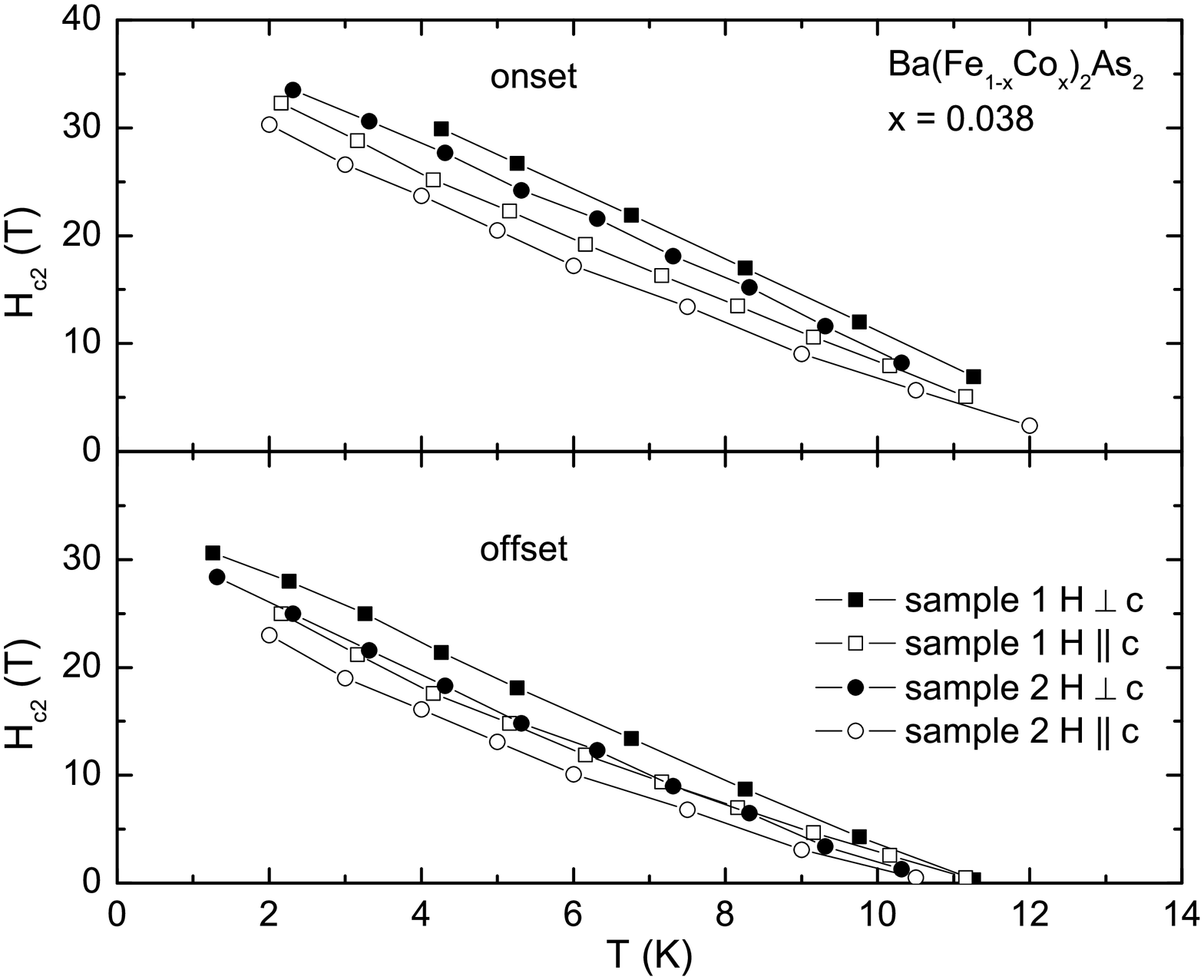}
\end{center}
\caption{Anisotropic $H_{c2}(T)$ curves determined for two single crystalline samples of $x = 0.038$ Ba(Fe$_{1-x}$Co$_x$)$_2$As$_2$ using onset criterion (upper panel) and offset criterion (lower panel).  $H \| c$ data shown as open symbols and $H \perp c$ data shown as filled symbols.}\label{F12}
\end{figure}

\clearpage

\begin{figure}
\begin{center}
\includegraphics[angle=0,width=120mm]{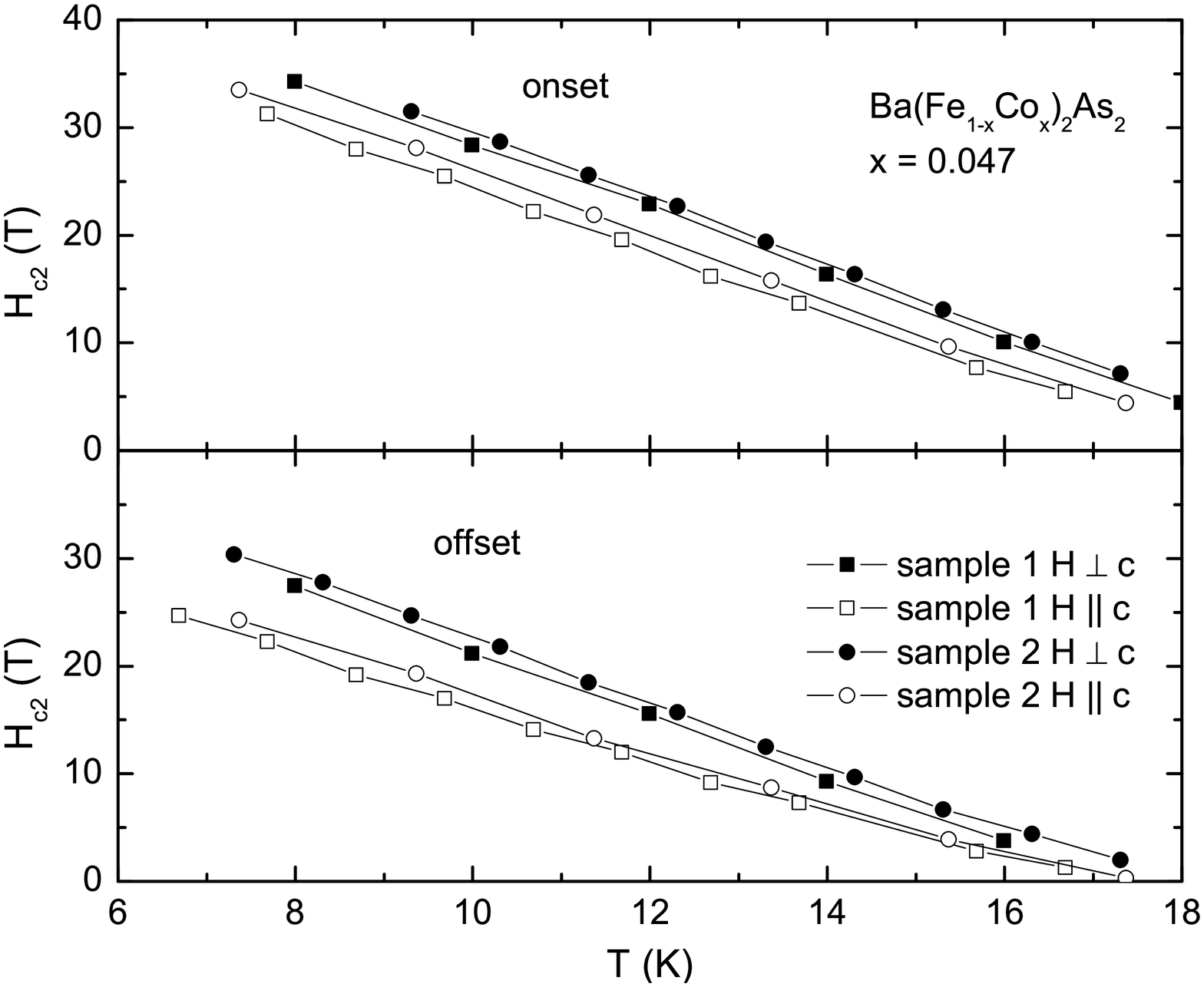}
\end{center}
\caption{Anisotropic $H_{c2}(T)$ curves determined for two single crystalline samples of $x = 0.047$ Ba(Fe$_{1-x}$Co$_x$)$_2$As$_2$ using onset criterion (upper panel) and offset criterion (lower panel).  $H \| c$ data shown as open symbols and $H \perp c$ data shown as filled symbols.}\label{F13}
\end{figure}

\clearpage

\begin{figure}
\begin{center}
\includegraphics[angle=0,width=120mm]{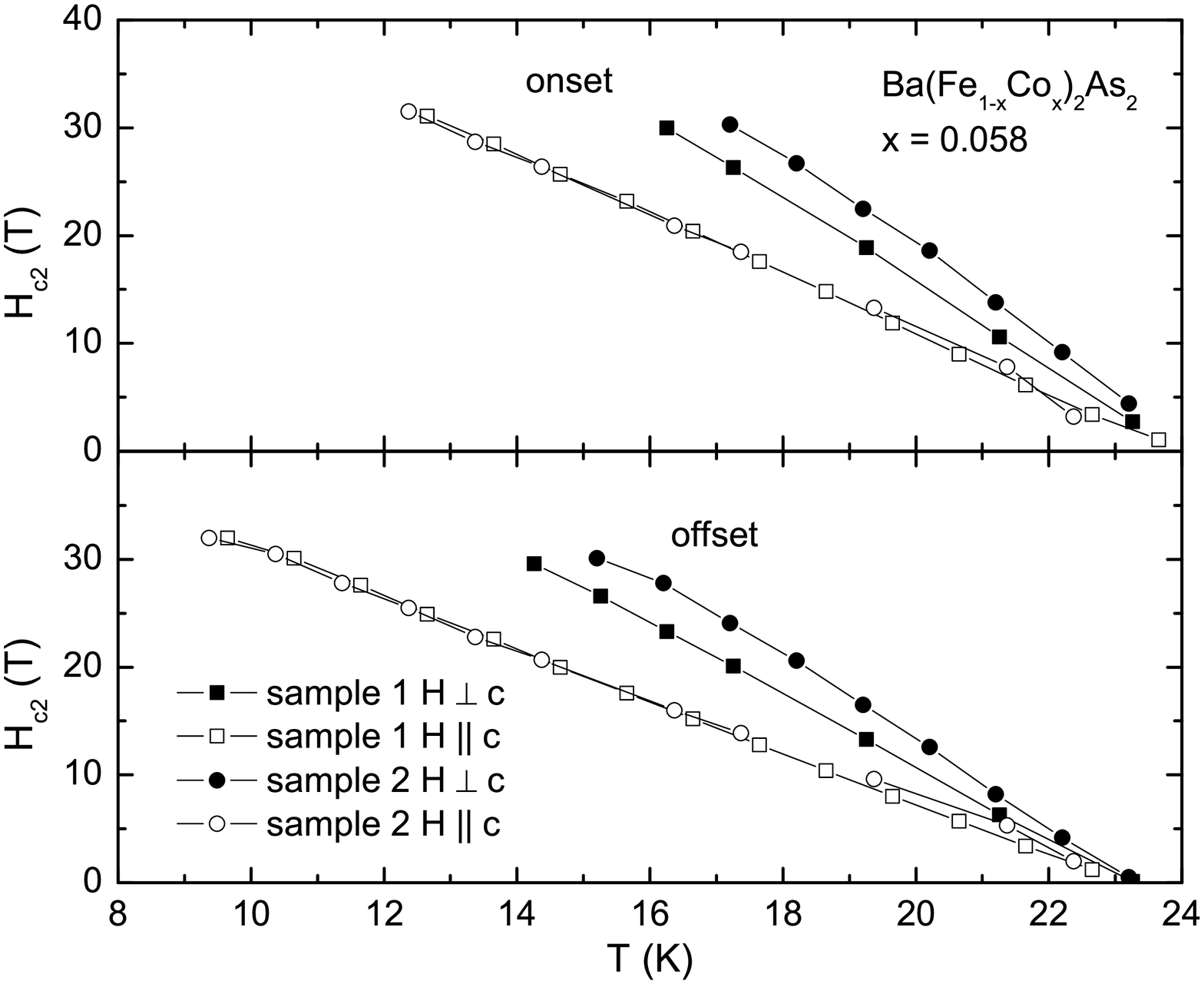}
\end{center}
\caption{Anisotropic $H_{c2}(T)$ curves determined for two single crystalline samples of $x = 0.058$ Ba(Fe$_{1-x}$Co$_x$)$_2$As$_2$ using onset criterion (upper panel) and offset criterion (lower panel).  $H \| c$ data shown as open symbols and $H \perp c$ data shown as filled symbols.}\label{F14}
\end{figure}

\clearpage

\begin{figure}
\begin{center}
\includegraphics[angle=0,width=120mm]{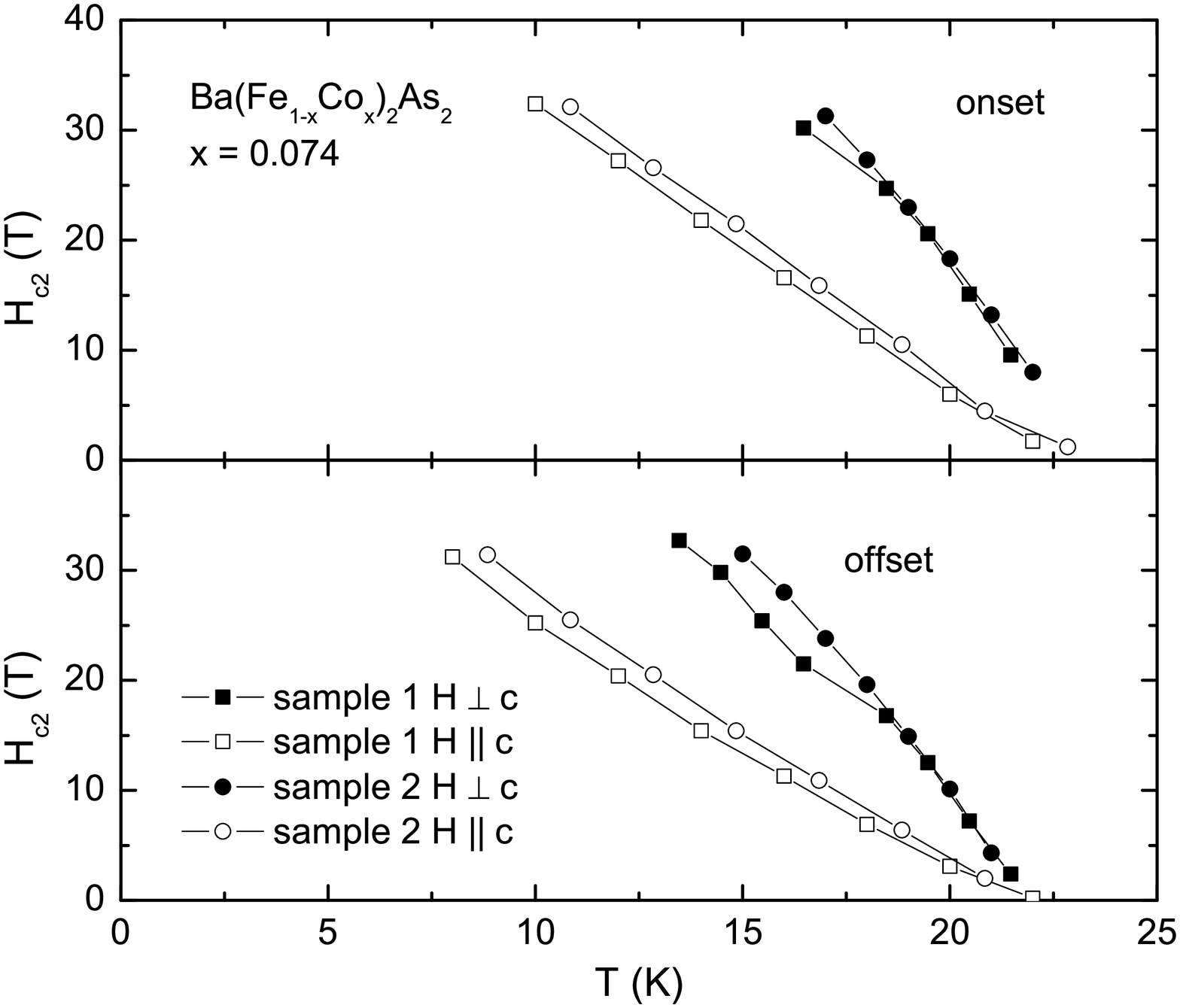}
\end{center}
\caption{Anisotropic $H_{c2}(T)$ curves determined for two single crystalline samples of $x = 0.074$ Ba(Fe$_{1-x}$Co$_x$)$_2$As$_2$ using onset criterion (upper panel) and offset criterion (lower panel).  $H \| c$ data shown as open symbols and $H \perp c$ data shown as filled symbols.}\label{F15}
\end{figure}

\clearpage

\begin{figure}
\begin{center}
\includegraphics[angle=0,width=120mm]{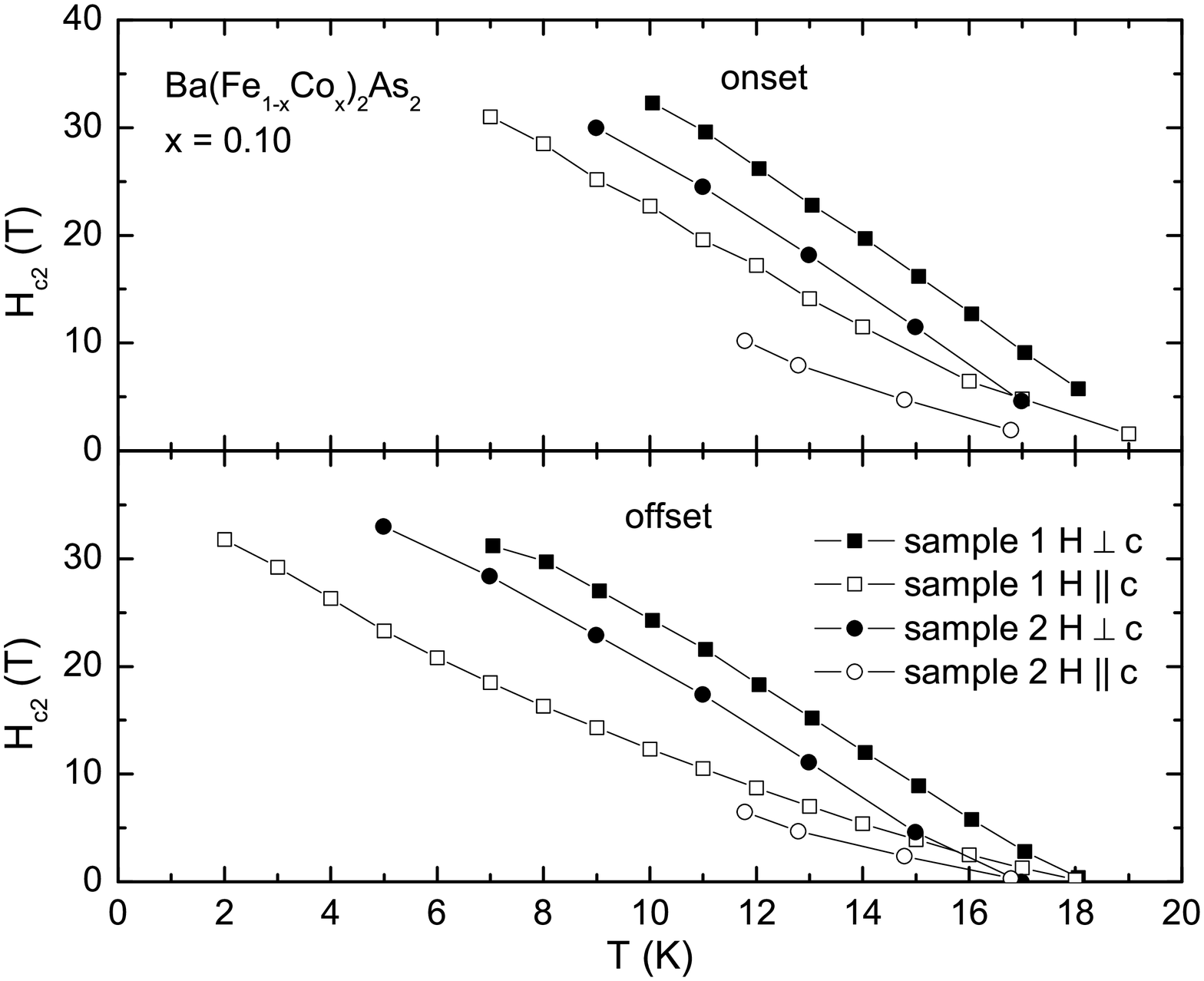}
\end{center}
\caption{Anisotropic $H_{c2}(T)$ curves determined for two single crystalline samples of $x = 0.100$ Ba(Fe$_{1-x}$Co$_x$)$_2$As$_2$ using onset criterion (upper panel) and offset criterion (lower panel).  $H \| c$ data shown as open symbols and $H \perp c$ data shown as filled symbols.}\label{F16}
\end{figure}

\clearpage

\begin{figure}
\begin{center}
\includegraphics[angle=0,width=120mm]{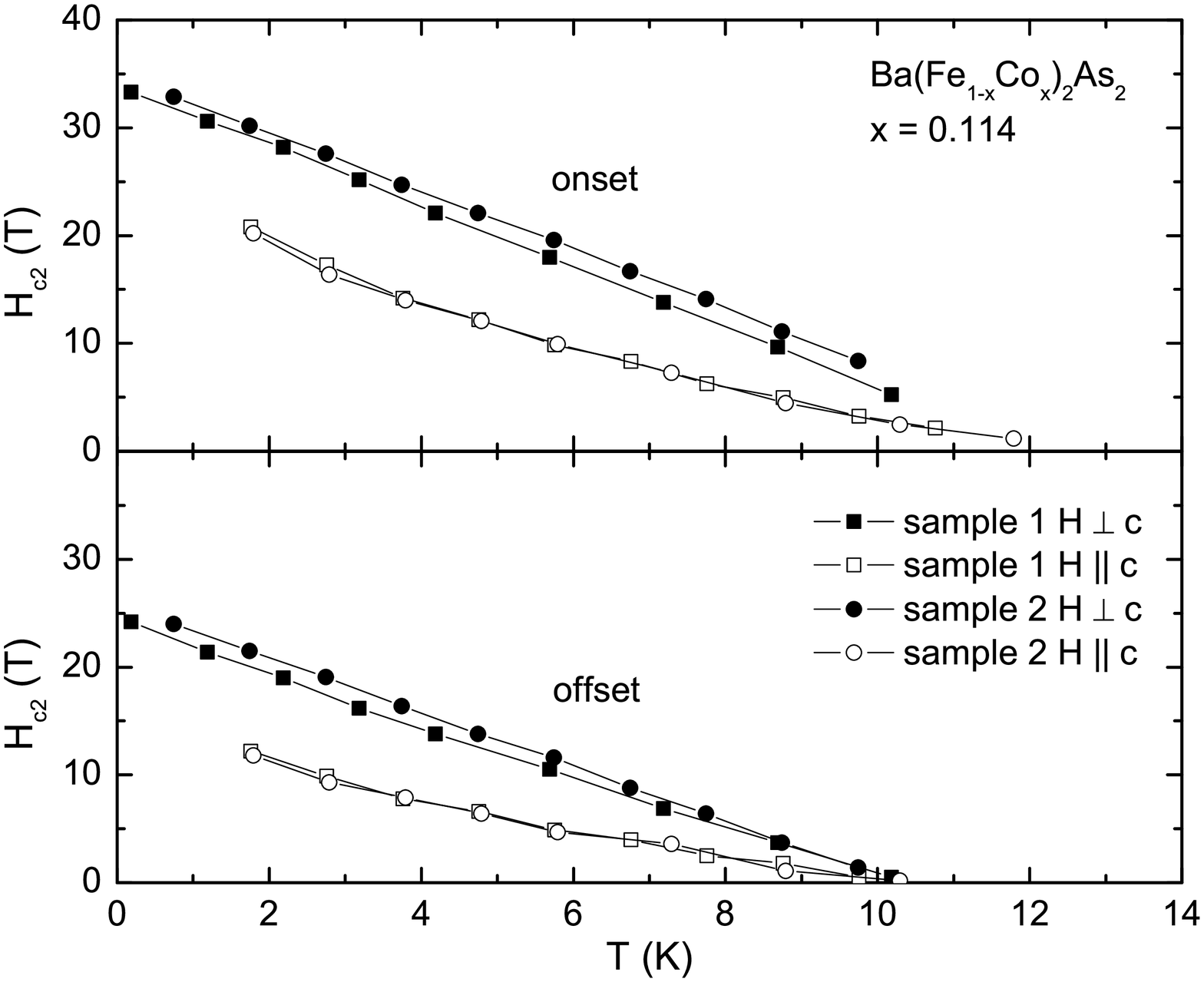}
\end{center}
\caption{Anisotropic $H_{c2}(T)$ curves determined for two single crystalline samples of $x = 0.114$ Ba(Fe$_{1-x}$Co$_x$)$_2$As$_2$ using onset criterion (upper panel) and offset criterion (lower panel).  $H \| c$ data shown as open symbols and $H \perp c$ data shown as filled symbols.}\label{F17}
\end{figure}

\clearpage

\begin{figure}
\begin{center}
\includegraphics[angle=0,width=120mm]{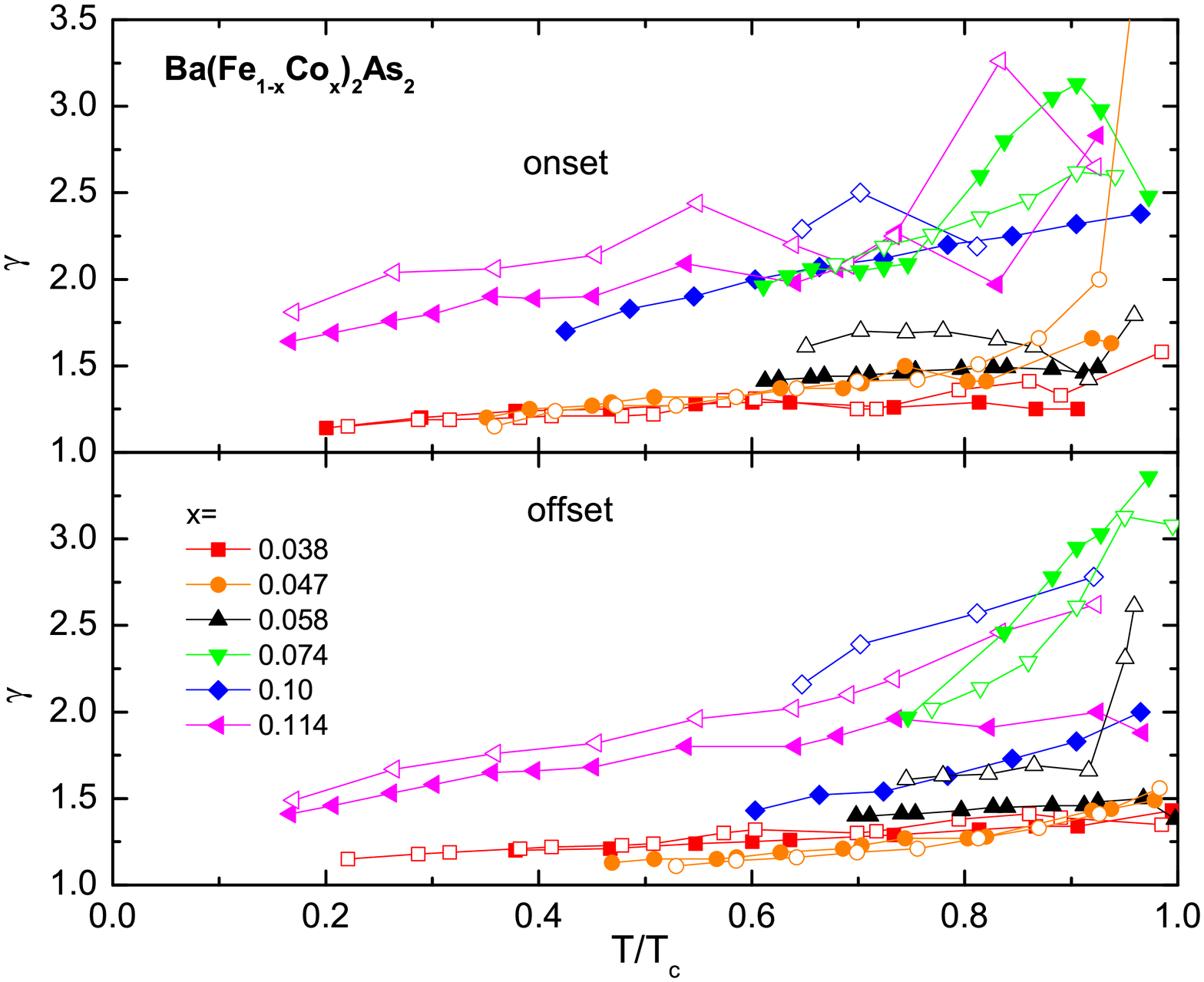}
\end{center}
\caption{(Color online) Anisotropy of the upper critical field, $\gamma = H_{c2}^{\perp c}(T)/H_{c2}^{\|c}(T)$, as a function of effective temperature, $T/T_c$, for Ba(Fe$_{1-x}$Co$_x$)$_2$As$_2$ single crystals.  Upper panel:  onset criterion; lower panel:  offset criterion. Open and filled symbols are the data for two samples with the same Co concentration.}\label{F18}
\end{figure}


\begin{thebibliography}{99}

\bibitem{kam08a}  Y. Kamihara, T. Watanabe, M. Hirano, H Hosono, Journal of the American Chemical Society  {\bf 130}, 3296 (2008).

\bibitem{rot08a} Marianne Rotter, Marcus Tegel, and Dirk Johrendt, Phys. Rev. Lett., {\bf 101}, 107006 (2008).

\bibitem{nin08a} N. Ni, S. L. Bud'ko, A. Kreyssig, S. Nandi, G. E. Rustan, A. I. Goldman, S. Gupta, J. D. Corbett, A. Kracher, and P. C. Canfield, Phys. Rev. B {\bf 78}, 014507 (2008).

\bibitem{wan08a} X. F. Wang, T. Wu, G. Wu, H. Chen, Y. L. Xie, J. J. Ying, Y. J. Yan, R. H. Liu and X. H. Chen, arXiv:0806.2452, unpublished.

\bibitem{yan08a} J.-Q. Yan, A. Kreyssig, S. Nandi, N. Ni, S. L. Bud'ko, A. Kracher, R. J. McQueeney, R. W. McCallum, T. A. Lograsso, A. I. Goldman, and P. C. Canfield, Phys. Rev. B {\bf 78}, 024516 (2008)

\bibitem{nin08b} N. Ni, S. Nandi, A. Kreyssig, A. I. Goldman, E. D. Mun, S. L. Bud'ko, and P. C. Canfield, Phys. Rev. B {\bf 78}, 014523 (2008)

\bibitem{ron08a} F. Ronning, T. Klimczuk, E.D. Bauer, H. Volz, J.D. Thompson, J. Phys.: Condens. Matter, {\bf 20}, 322201 (2008).

\bibitem{wug08a} G. Wu, H. Chen, T. Wu, Y. L. Xie, Y. J. Yan, R. H. Liu, X. F. Wang, J. J. Ying and X. H. Chen, J. Phys.: Condens. Matter, {\bf 20}, 422201 (2008).

\bibitem{sef08a} Athena S. Sefat, Rongying Jin, Michael A. McGuire, Brian C. Sales, David J. Singh, and David Mandrus, Phys. Rev. Lett., {\bf 101}, 117004 (2008).

\bibitem{lei08a} A. Leithe-Jasper, W. Schnelle, C. Geibel, and H. Rosner, arXiv:0807.2223, unpublished.

\bibitem{ahi08a} K. Ahilan, J. Balasubramaniam, F. L. Ning, T. Imai, A. S. Sefat, R. Jin, M. A. McGuire, B. C. Sales, and D. Mandrus, J. Phys.: Condens. Matter {\bf 20}, 472201 (2008).

\bibitem{yam08a}A. Yamamoto, J. Jaroszynski, C. Tarantini, L. Balicas, J. Jiang, A. Gurevich,  D. C. Larbalestier, R. Jin, A.S. Sefat, M.A. McGuire, B.C. Sales, D.K. Christen, and D. Mandrus, arXiv:0810.0699, unpublished.

\bibitem{pro08a} R. Prozorov, N. Ni, M. A. Tanatar, V. G. Kogan, R. T. Gordon, C. Martin, E. C. Blomberg, P. Prommapan, J. Q. Yan, S. L. Bud'ko, and P. C. Canfield, arXiv:0810.1338, unpublished.

\bibitem{gor08a} R. T. Gordon, N. Ni, C. Martin, M. A. Tanatar, M. D. Vannette, H. Kim, G. Samolyuk, J. Schmalian, S. Nandi, A. Kreyssig, A. I. Goldman, J. Q. Yan, S. L. Bud'ko, P. C. Canfield, and R. Prozorov, arXiv:0810.2295, unpublished.

\bibitem{can92a}  P. C. Canfield, Z.  Fisk, Phil. Mag. B  {\bf 65}, 1117 (1992).

\bibitem{hol97a} T. J. B. Holland, and S. A. T. Redfern, Mineral. Mag. {\bf 61}, 65 (1997).

\bibitem{rot08b} M. Rotter, M. Tegel, D. Johrendt, I. Schellenberg, W. Hermes, and R. Pöttgen, Phys. Rev. B {\bf 78}, 020503, (2008).

\bibitem{hua08a} Q. Huang, Y. Qiu, Wei Bao, M.A. Green, J.W. Lynn, Y.C. Gasparovic, T. Wu, G. Wu, X. H. Chen, arXiv:0806.2776, unpublished.

\bibitem{tor08a} M. S. Torikachvili, S. L. Bud'ko, N. Ni, and P. C. Canfield, Phys. Rev. B {\bf 78}, 104527 (2008).

\bibitem{mar08a} C. Martin, R. T. Gordon, M. A. Tanatar, M. D. Vannette, M. E. Tillman, E. D. Mun, P. C. Canfield, V. G. Kogan, G. D. Samolyuk, J. Schmalian, and R. Prozorov, arXiv:0807.0876, unpublished.

\bibitem{wel08a} U. Welp, R. Xie, A. E. Koshelev, W. K. Kwok, P. Cheng, L. Fang, and H.-H. Wen, arXiv:0807.4196, unpublished.

\bibitem{bal08a} L. Balicas, A. Gurevich, Y. J. Jo, J. Jaroszynski, D. C. Larbalestier, R. H. Liu, H. Chen, X. H. Chen, N. D. Zhigadlo, S. Katrych, Z. Bukowski, and J. Karpinski, arXiv:0809.4223, unpublished.

\bibitem{jar08a} J. Jaroszynski, F. Hunte, L. Balicas, Youn-jung Jo, I. Rai\v{c}evi\'{c}, A. Gurevich, D. C. Larbalestier, F. F. Balakirev, L. Fang, P. Cheng, Y. Jia and H. H. WenarXiv:0810.2469, unpublished.

\bibitem{wer66a} N. R. Werthamer, E. Helfand, and P. C. Hohenberg, Phys. Rev. {\bf 147}, 295 (1966).

\bibitem{shu98a} S. V. Shulga, S.-L. Drechsler, G. Fuchs, K.-H. Müller, K. Winzer, M. Heinecke, and K. Krug, Phys. Rev. Lett., {\bf 80}, 1730 (1998).

\bibitem{alt08a} M. Altarawneh, K. Collar, C. H. Mielke, N. Ni, S. L. Bud'ko, P.C. Canfield, arXiv:0807.4488, unpublished.

\bibitem{yua08a} H. Q. Yuan, J. Singleton, F. F. Balakirev, G. F. Chen, J. L. Luo, N. L. Wang, arXiv:0807.3137, unpublished.

\bibitem{rot08c} Marianne Rotter, Michael Pangerl, Markus Tegel, and Dirk Johrendt, Angew. Chem. Int. Ed. {\bf 47} 7949 (2008).

\bibitem{che08a} H. Chen, Y. Ren, Y. Qiu, Wei Bao, R. H. Liu, G. Wu, T. Wu, Y. L. Xie, X. F. Wang, Q. Huang, and X. H. Chen, arXiv:0807.3950, unpublished.
    
\bibitem{gol08a} A. I. Goldman, D. N. Argyriou, B. Ouladdiaf, T. Chatterji, A. Kreyssig, S. Nandi, N. Ni, S. L. Bud'ko, P. C. Canfield, and R. J. McQueeney, Phys. Rev. B {\bf 78}, 100506 (2008).


\end{thebibliography}
\end{document}